\documentclass{aa} 
\usepackage{graphicx}
\usepackage{natbib}
\bibpunct{(}{)}{;}{a}{}{,}

\begin{document}

\title{Multi-order TODCOR: application to observations taken
with the CORALIE echelle spectrograph}

\subtitle{I. The system \object{HD\,41004}\thanks{Based on
    observations collected at the La Silla Observatory, ESO (Chile),
    with the CORALIE spectrograph at the 1.2-m Euler Swiss telescope}
    \fnmsep \thanks{Table 4  is only available in electronic form at
    the CDS via anonymous ftp to cdsarc.u-strasbrg.fr (130.79.128.5)
    or via http ://cdsweb.u-strasbrg.fr/A+A.htx}}

\author{S. Zucker\inst{1,2}    \and 
        T. Mazeh\inst{1}     \and 
        N.~C. Santos\inst{3} \and
        S. Udry\inst{3}      \and
        M. Mayor\inst{3}
}

\institute{School of Physics and Astronomy, Raymond and Beverly
  Sackler Faculty of Exact Sciences, Tel Aviv University, Tel Aviv,
  Israel 
\and 
Present address: Dept. of Geophysics and Planetary
  Sciences, Raymond and Beverly Sackler Faculty of Exact Sciences, Tel
  Aviv University, Tel Aviv, Israel 
\and 
Observatoire de Gen\`{e}ve,
  51 ch.\ des Maillettes, CH-1290 Sauverny, Switzerland}

\offprints{Shay Zucker, \email{shay@wise.tau.ac.il}}

\date{Received / Accepted}

\abstract{ This paper presents an application of the TwO-Dimensional
  CORrelation (TODCOR) algorithm to multi-order spectra. The
  combination of many orders enables the detection and measurement of
  the radial velocities of very faint companions. The technique is
  first applied here to the case of \object{HD\,41004}, where the
  secondary is $3.68$ magnitudes fainter than the primary in the
  V~band. When applied to CORALIE spectra of this system, the
  technique measures the secondary velocities with a precision of
  $0.6\,\mathrm{km\,s}^{-1}$ and facilitates an orbital solution of
  the \object{HD\,41004\,B} subsystem. The orbit of
  \object{HD\,41004\,B} is nearly circular, with a companion of a
  $19\,M_\mathrm{J}$ minimum mass.  The precision achieved for the
  primary is $10\,\mathrm{m\,s}^{-1}$, allowing the measurement of a
  long-term trend in the velocities of \object{HD\,41004\,A}.
  
  \keywords{ methods: data analysis -- techniques: radial velocities
    -- binaries: spectroscopic -- stars: individual:
    \object{HD\,41004} -- stars: low-mass, brown dwarfs } }

\titlerunning{Multi-order TODCOR: application to CORALIE \& ELODIE}

\maketitle

\section{Introduction}

TODCOR (TwO-Dimensional CORrelation) is a two-dimensional correlation
technique to derive the radial velocities of both components of
double-lined binary spectra \citep{ZucMaz1994}. It was introduced as a
generalization of the (one-dimensional) cross-correlation technique
\citep{Sim1974, TonDav1979} to deal with the difficulties encountered
in double-lined spectra, when the lines of the two components cannot
be easily resolved. Assuming the observed spectrum is a combination of
two known spectra shifted by the radial velocities of the two
components, TODCOR calculates the correlation of the observed spectrum
against a combination of two templates with different shifts.  The
result is a two-dimensional correlation function, whose peak
simultaneously identifies the radial velocities of both the primary
and the secondary.

One of the advantages of TODCOR is its ability to use different
templates for the primary and the secondary. When the primary and the
secondary are of different spectral types, the simultaneous use of two
different templates utilizes all the spectral information contained in
the observed spectrum of the combined system. This feature of TODCOR
is most important when deriving the radial velocities of faint
secondaries \citep{MazZuc1994}.  However, the ability to detect faint
secondaries using TODCOR is still limited, depending mainly on the
signal-to-noise ratio and the number of spectral lines. Therefore,
\citet{Mazetal2002} and \citet{Praetal2002}, who searched for faint
companions, had to apply TODCOR on IR spectra, where the flux ratio is
favorable for detecting cooler stars.

The modern spectrographs offer another path to enhance the ability to
detect faint companions.  Due to the progress in detector technology,
many of the modern spectrographs produce multi-order spectra. In order
to enhance our ability to detect a faint companion, while maximizing
the precision of the measured velocities, we need to combine the
spectral information in all the relevant orders.  Originally, TODCOR
was devised to analyse only single-order spectra, and further
generalization is therefore needed in order to use the information in
multiple orders.

When applying TODCOR to each order separately, the weak signal of the
secondary may produce a certain local peak of the correlation function
at the correct secondary velocity. However, this peak can be easily
topped by spurious random peaks.  This rules out, for example,
calculation of the radial velocity independently for each order and
then averaging the velocities, since many of the secondary velocities
would have resulted from wrong peaks. Concatenation of the spectral
orders to one single spectrum would require special treatment to the
gaps and overlaps between adjacent orders. Co-adding overlapping
regions and bridging the gaps both require interpolation, which would
introduce artificial noise into the analysed spectrum.

The approach we suggest here is to calculate the correlation function
for each order separately, and then combine the correlation functions
of all the orders. The combination emphasizes the relevant correlation
peak, and averages out the spurious random peaks. A simple average of
the correlation functions may not be efficient enough, since the
combination scheme has to consider the different spectral information
in the different orders, and weigh them accordingly.  \citet{Zuc2003}
introduces such a scheme, based on a few plausible statistical assumptions.

Using this scheme, multi-order TODCOR is applied in this work to the
CORALIE spectra of \object{HD\,41004}.  As \citet{Sanetal2002} have
shown, these spectra are composed of the spectra of the two visual
components of the system, that are separated by $0.5\arcsec$.
\citeauthor{Sanetal2002} found that the radial velocities measured by
CORALIE for these spectra showed a minute periodical variability.
They suggested that this apparent variation was actually related to a
much larger variation in the velocity of the faint B component.  The
presence of the spectrum of \object{HD\,41004\,B} caused a variable
asymmetry in the line shape of the composite spectra, which was
reflected in a minute variability of the measured radial velocity.

TODCOR, when applied to the multi-order CORALIE spectra of
\object{HD\,41004}, derived the radial velocities of {\it both}
components without any assumptions regarding the orbit. The derived
velocities confirm the conjecture of \citeauthor{Sanetal2002}: the B
component indeed moves periodically with a $1.3$-day period, and the A
component shows a long-term trend.  We present here an orbital
solution for the B component and derive the slope of the long-term
trend of the A component, based on the precise velocities derived by
TODCOR.

The next section briefly reviews the previous results concerning
\object{HD\,41004}.  The analysis and its results are presented in
Sections~\ref{analsect} and~\ref{ressect}. Section~\ref{discsect}
discusses the results and their implications. 
Section~\ref{concsect} concludes the paper with a few remarks.

\section{Characteristics of \object{HD\,41004}}
\label{stelcharsect}

\object{HD\,41004} is a visual double system, consisting of a K1V--M2V
pair. According to the Hipparcos catalogue, the pair is a
common-proper-motion pair, with a V-magnitude difference of $3.68$ and
a separation of $0.541\arcsec\pm0.033\arcsec$. The basic stellar
parameters of \object{HD\,41004\,A} are summarized in
Table~\ref{stelchar}, reproducing Table~1 of \citet{Sanetal2002}.  The
effective temperature, surface gravity and metallicity were calculated
by \citeauthor{Sanetal2002} through Str\"{o}mgren photometry, which
also showed that the star is photometrically stable within the
instrumental precision.  An independent estimate of the metallicity
was obtained from analysis of the Cross-Correlation Function of
CORALIE. The two independent estimates of the metallicity are quoted
in the table.

\begin{table}
\caption{Stellar parameters of \object{HD\,41004\,A} (reproduced from
 \citealt{Sanetal2002})}
\begin{tabular}[]{ll}
Parameter                 & Value            \\
\noalign{\vspace{0.05cm}}
\hline
\noalign{\vspace{0.025cm}}
\hline
\noalign{\vspace{0.05cm}}
Spectral Type              & K1V/K2V          \\
Parallax [mas]             & $23.24$          \\
Distance [pc]              & $43$             \\
$m_v$                      & $8.65$           \\
$B-V$                      & $0.887$          \\
$T_\mathrm{eff}$ [K]       & $5010$           \\
$\log g$ [cgs]             & $4.42$           \\
$M_V$                      & $5.48$           \\
Luminosity [$L_{\sun}$]    & $0.65$           \\
Mass [$M_{\sun}$]          & $\sim0.7$        \\
$\log R'_\mathrm{HK}$      & $-4.66$          \\
Age [Gyr]                  & $1.6$            \\
$P_\mathrm{rot}$ [days]    & $\sim27$         \\
$v \sin i$ [km\,s$^{-1}$]  & $1.22$           \\
$[\mathrm{Fe}/\mathrm{H}]$ & $-0.09$/$+0.10$    
\end{tabular}
\label{stelchar}
\end{table}

The principal result of \citet{Sanetal2002} is the detection of a
radial-velocity periodical variability in the CORALIE spectra of
\object{HD\,41004}. The variability pattern presented in their paper
is consistent with the presence of a planet orbiting
\object{HD\,41004\,A} with a $1.3$-day period. However,
\citeauthor{Sanetal2002} rejected this possibility based on the
bisector shape analysis of the CORALIE spectra. This analysis revealed
a periodic line shape variation, having the same period as the
radial-velocity variation.

The interpretation \citeauthor{Sanetal2002} suggested is the presence
of an object orbiting the M2V star \object{HD\,41004\,Ba}.  In this
model A (the K star) and B are orbiting each other in a wide orbit,
while B is further composed of two objects, Ba (the M star) and Bb, in
a very close orbit, with a period of $1.3$ days.  According to this
interpretation, the CORALIE spectra of \object{HD\,41004} are a
combination of the A and B components, because their separation is
much smaller than the diameter of the CORALIE fiber. Since the flux of
the B component in the relevant wavelength range is only about $3\%$
of the flux of the primary, large radial-velocity variations of B
cause only small variations of the measured radial velocity, but their
effect is manifested in the bisector shape.

\citeauthor{Sanetal2002} performed some simulations to test their
interpretation and concluded that their results are consistent with
the presence of a brown dwarf orbiting \object{HD\,41004\,Ba}, with a
radial-velocity amplitude of the order of $5\,\mathrm{km\,s}^{-1}$.
They also found a long-term linear trend in the radial velocities,
which did not agree with the older measurements. They ascribed it to
the motion of \object{HD\,41004\,A} around \object{HD\,41004\,B}, but
did not rule out an additional component that may be involved in this
variation.

\section{Analysis}
\label{analsect}

The data we analysed comprised the 86 CORALIE spectra of
\object{HD\,41004} used by \citet{Sanetal2002}, which were obtained
between November 2001 and February 2002, except for one spectrum
obtained in December 2000.  The system was further monitored and in
this paper we add 17 spectra obtained as of March 2002.

CORALIE is a fiber-fed, cross-dispersed echelle spectrograph, mounted
on the $1.2$-m Leonard Euler telescope at La Silla
\citep{Queetal1999}.  With a resolution of
$\lambda/\Delta\lambda=50\,000$, it covers the wavelength range
$3800$--$6900$\,\AA, with $68$ echelle orders. The CORALIE system uses
a software code that produces automatically the radial velocities, but
for this work we used the reduced spectral orders.

The analysis used $32$ CORALIE orders within the spectral range
$4780$--$6820$\,\AA, after having excluded the orders that are heavily
polluted by telluric lines. Bluer orders were excluded because the
secondary signal is expected to be too weak in these orders.
Multi-order TODCOR was used to combine the correlation functions from
the $32$ orders. Note that the pixel-to-radial-velocity scale is
different for each order.  Therefore, before combining the correlation
functions we had to interpolate and re-sample them into a
pre-determined scale.

TODCOR requires two spectral templates as similar as possible to the
expected primary and secondary spectra, and an assumed value for the
flux ratio between the two spectra.  We searched for the best
templates in two datasets of spectra obtained by ELODIE and CORALIE.
The CORALIE templates, appropriate for solar-type stars, were obtained
as part of a program to derive precise abundances of planet-hosting
and non-planet-hosting stars \citep{Sanetal2000,Sanetal2001}. The
templates for the later stars were obtained using ELODIE as part of a
program studying the binarity of close M dwarfs
\citep{Deletal1998,Deletal1999}.  The flux-ratio information was taken
from \citet{Pic1998}, using his measurements of typical stellar
spectral energy distribution (SED), normalized according to the
V-magnitude difference of \object{HD\,41004}. By convolving the
templates with a rotational broadening profile
\citep[e.g.,][]{Gra1976}, we created an additional degree of freedom which
expanded our template library, allowing a better fit of the templates
to the observed spectra.

In all the configurations of two templates that had spectral types
similar to K1V and M2V, the secondary radial velocities exhibited a
clear nearly circular orbit with an amplitude of around
$6\,\mathrm{km\,s}^{-1}$, and a linear trend of the primary. We
finally chose the configuration which yielded the smallest residuals
in the secondary velocities, relative to the best-fit orbital
solution.  Table~\ref{temptable} lists the known stellar
characteristics of \object{HD\,52698} and \object{GJ\,393} -- the two
templates finally chosen.  
The stellar parameters of \object{HD\,52698} (effective temperature, surface
gravity and metallicity) were derived as in \citet{Sanetal2001},
whereas an estimate of the $v \sin i$ was obtained using the calibration of
the CORALIE Cross-Correlation Function width (see the Appendix in 
\citet{Sanetal2002}).

Table~\ref{temptable} quotes the original equatorial rotational
velocities for both templates.  Even mild broadening of the primary
template ($v \sin i = 1\,\mathrm{km\,s}^{-1}$), on top of its original
$v\sin i$ of $2\,\mathrm{km\,s}^{-1}$, degraded the solution
considerably.  On the other hand, the secondary template could be
broadened by $v \sin i$ up to $5\,\mathrm{km\,s}^{-1}$ without
significantly affecting the solution. Finally we used a broadening of
$v \sin i = 1\,\mathrm{km\,s}^{-1}$, which yielded the smallest
residuals of the secondary velocities.

\begin{figure}
\resizebox{\hsize}{!}{\includegraphics{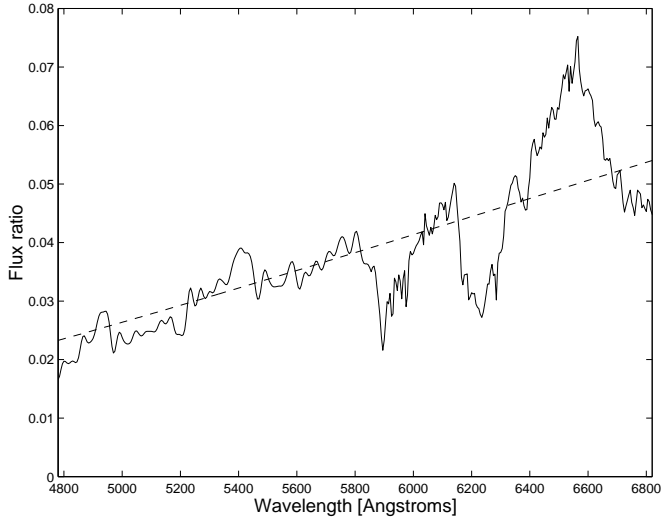}}
\caption{ 
  The flux ratio in the relevant wavelength range, calculated
  according to \citet{Pic1998}. The dashed line represents the flux
  ratio corresponding to black-body radiation laws of 3500\,K and
  5240\,K.}
\label{alpha}
\end{figure}

Figure~\ref{alpha} shows the flux ratio as a function of wavelength,
corresponding to the chosen spectral types of K1V and M2.5V, using the
SED library of \citet{Pic1998}. The dashed line in the figure shows
the flux ratio obtained by assuming a black-body radiation law for the
two spectra, with a temperature of 3500\,K for the secondary.  It is
clear that while the black-body model roughly fits the detailed
flux-ratio, it is not accurate enough and there are large differences,
e.g., around $6500$\,\AA.  In any case, using a constant flux ratio as
in the single-order application of TODCOR is clearly not sufficient
for such a wide spectral range.

\begin{table}
\caption{Known stellar characteristics of the templates finally chosen}
\begin{tabular}[]{lll}
                           & Primary              & Secondary              \\
Parameter                  & Template             & Template               \\
\noalign{\vspace{0.05cm}}
\hline
\noalign{\vspace{0.025cm}}
\hline
\noalign{\vspace{0.05cm}}
Name                               & \object{HD\,52698}     & \object{GJ\,393}       \\
Spectral Type                      & K1\footnotemark[1]     & M2.5\footnotemark[1]   \\
$B-V$                              & $0.89$\footnotemark[1] & 1.52\footnotemark[1]   \\
$T_\mathrm{eff}$ [K]               & $5235$                 & N/A                    \\
$\log g$                           & $4.69$                 & N/A                    \\
${[\mathrm{Fe}/\mathrm{H}]}$       & $0.21$                 & N/A                    \\
$v \sin i$ [km\,s$^{-1}$]          & $2.0$                  & $<2.9$\footnotemark[2] \\
\\
\noalign{\vspace{0.05cm}}
\hline
\noalign{\vspace{0.05cm}}
\end{tabular}
\\
1. \citealt{Hunetal1999} \\
2. \citealt{Deletal1998}
\label{temptable}
\end{table}

\begin{figure}
\centering
\includegraphics[width=9cm]{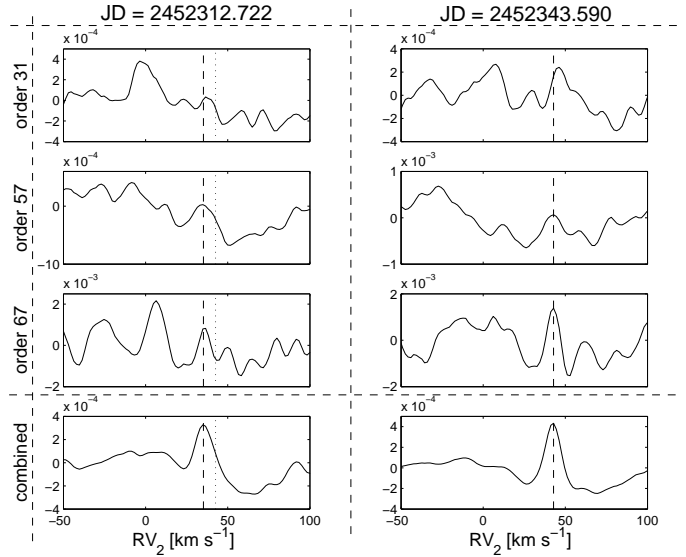}
\caption{ 
  The upper three panels in each column show ``cuts'' (see text) of
  the two-dimensional correlation function for three selected orders.
  The dashed lines represent the best-fit secondary velocities while
  the dotted lines represent the primary velocity.  It is not shown on
  the right column where it almost coincides with the secondary
  velocity. The lower panel in each column shows the ``cut'' of the
  function obtained after combining the corresponding correlation
  functions of all the 32 orders. A third-order best-fit polynomial
  was subtracted from all the shown functions to accentuate the local
  peak.}
\label{demo}
\end{figure}

Figure~\ref{demo} demonstrates the way multi-order TODCOR measures the
radial velocity of \object{HD\,41004\,B}.  The Figure presents
one-dimensional ``cuts'' of the two-dimensional correlation function.
In each ``cut'' the correlation is shown as a function of the
secondary velocity, while the primary velocity is fixed according to
the location of the two-dimensional maximum.  Because of the small
flux-ratio between the two templates, we expect the correlation value
to change very little when the secondary template alone is shifted.
However, in each ``cut'' we still expect to see a relative peak around
the correct secondary velocity.  The Figure is divided into two
columns, corresponding to two different exposures
($\mathrm{JD}=2\,452\,312.722$ and $\mathrm{JD}=2\,452\,343.590$).
Using the best-fit orbit, the expected secondary velocity for the
spectrum used in the left column of the Figure is
$35.2\,\mathrm{km\,s}^{-1}$, while for the right column it is
$42.2\,\mathrm{km\,s}^{-1}$. These velocities are marked on the figure
by dashed lines. The primary velocity is represented by a dotted line
in the left column, while in the right column it almost coincides with
the secondary velocity. For this graphic demonstration only, a
third-order best-fit polynomial was subtracted from all the plotted
functions in order to accentuate the local peak.

The three upper panels in each column demonstrate the problems in the
single-order TODCOR. On the left, the first upper panel exhibits a
very prominent peak in a wrong velocity. The second and third panels
show a moderate peak at about the expected velocity In almost all $32$
orders, some local peaks appeared around the correct velocity.  The
lower panel shows the result of combining the correlation functions of
all the analysed orders. The correct peak is clearly emphasized
relative to the spurious peaks.

Comparing the two columns of Figure~\ref{demo}, we see that we cannot
know in advance which orders present the correct peak. Thus, we have
to combine all $32$ orders in order to have the correct peak
emphasized. The right column of the Figure demonstrates another
advantage of TODCOR: in this spectrum the secondary velocity is almost
identical to the primary velocity.  In the conventional
one-dimensional cross-correlation, there is no way to measure the two
velocities, due to the blending of the two correlation peaks.  TODCOR,
by using two different templates, allows the measurement of both
velocities.

The listing of the radial velocities of A and B and the
corresponding times, can be obtained at the CDS.  

\section{Results}
\label{ressect}

\begin{figure}
\resizebox{\hsize}{!}{\includegraphics{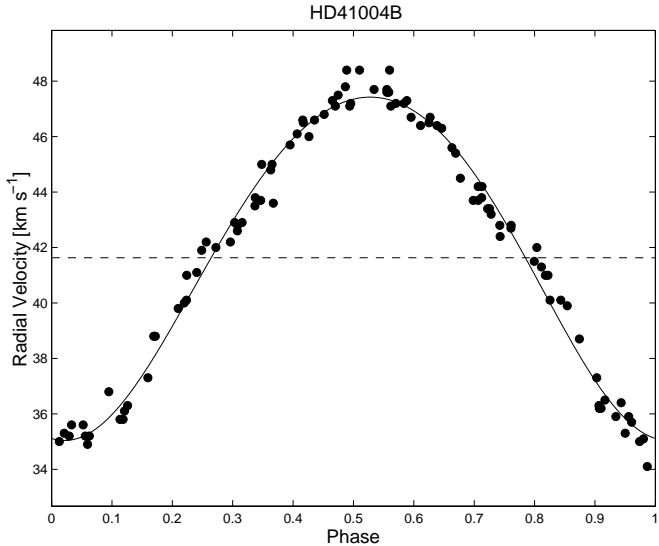}}
\caption{
Phased radial velocities of \object{HD\,41004\,B}} 
\label{orbit}
\end{figure}

Figure~\ref{orbit} shows the resulting orbit of \object{HD\,41004\,B},
while the orbital elements are summarized in Table~\ref{orbtable}.
Applying multi-order TODCOR to many orders ($32$), combined with the
large number of measurements ($103$) yielded a very precise orbital
solution. Thus, the radial-velocity amplitude ($K$) was found with a
precision of about $1\%$. This fine precision allowed also a very
accurate estimate of the orbital eccentricity ($e$). Although very
small, the eccentricity is still non-vanishing, with a significance
level of $2\cdot10^{-5}$ according to the \citet{LucSwe1971} test.
Assuming a mass of $\sim0.4\,M_{\sun}$ for \object{HD\,41004\,Ba}, the
companion minimum mass is $19\,M_\mathrm{J}$. The uncertainty of
$0.25\,M_\mathrm{J}$ does not take into account the uncertainty in the
mass of \object{HD\,41004\,Ba} -- $M_\mathrm{Ba}$. A $20\%$
uncertainty in $M_\mathrm{Ba}$ would result in a $2.3\,M_\mathrm{J}$
uncertainty in $M_\mathrm{Bb,min}$.

\begin{table}
\caption{Best-fit orbital solution of \object{HD\,41004\,B}}
\begin{tabular}[]{lll@{  \,$\pm$\,  }l}
\noalign{\vspace{0.05cm}}
\hline
\noalign{\vspace{0.025cm}}
\hline
\noalign{\vspace{0.05cm}}
$P$                          & [days]                 & $1.328199$        & $0.000081$   \\
$T$                          & [JD]                   & $2\,452\,339.212$ & $0.040$      \\
$e$                          &                        & $0.065$           & $0.014$      \\
$\gamma$                     & [km\,s$^{-1}$]         & $41.631$          & $0.057$      \\
$\omega$                     & [$\degr$]              & $171$             & $11$         \\ 
$K$                          & [km\,s$^{-1}$]         & $6.192$           & $0.081$      \\
$a \sin i$                   & [$10^{-3}$\,AU]        & $0.7544$          & $0.0098$     \\
$f(m)$                       & [$10^{-6}\,M_{\sun}$]  & $32.5$            & $1.3$        \\
$M_\mathrm{Bb,min}^\dagger$  & [$M_\mathrm{J}$]       & $18.64$           & $0.26$       \\
\noalign{\vspace{0.05cm}}
\hline
\noalign{\vspace{0.05cm}}
$N$                &                     & \multicolumn{2}{c}{103}  \\
$\sigma_\mathrm{O-C}$ & [km\,s$^{-1}$]   & \multicolumn{2}{c}{0.56} \\
\noalign{\vspace{0.05cm}}
\hline
\noalign{\vspace{0.05cm}}
\end{tabular}
\\
$^\dagger$Assuming $M_\mathrm{Ba} = 0.4\,M_{\sun}$.
\label{orbtable}
\end{table}

\begin{figure}
  \resizebox{\hsize}{!}{\includegraphics{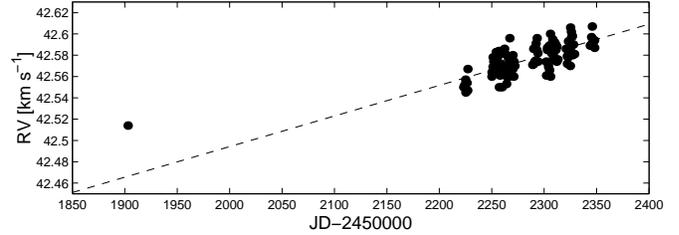}}
\caption{ Radial velocities of \object{HD\,41004\,A}. The dashed line
is the best fit to the velocities, ignoring the first isolated
velocity.}
\label{trend}
\end{figure}

\begin{figure}
\resizebox{\hsize}{!}{\includegraphics{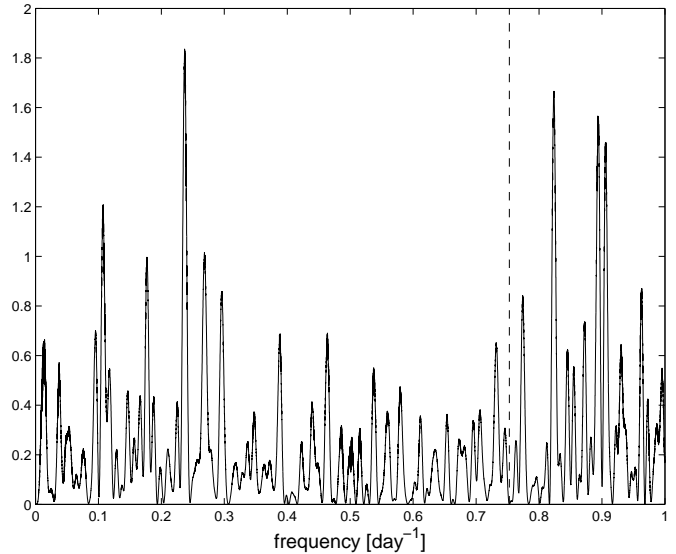}}
\caption{ Lomb-Scargle periodogram of the velocities of
  \object{HD\,41004\,A} after subtracting the linear trend. The dashed
  line marks the frequency corresponding to the $1.3$-day period of
  \object{HD\,41004\,B}.}
\label{periodogram}
\end{figure}

Figure~\ref{trend} shows the velocities of \object{HD\,41004\,A}.  The
velocities measured after $\mathrm{JD}=2,452,200$ show a clear linear
trend. The best-fit line has a slope of
$+105\,\pm\,10\,\mathrm{m\,s}^{-1}\,\mathrm{year}^{-1}$, and is also
shown in the Figure.  As \citet{Sanetal2002} have already noticed, the
first isolated measurement does not agree with the linear trend
implied by the other measurements.  A Lomb-Scargle periodogram of the
de-trended velocities (Figure~\ref{periodogram}) shows no hint of the
$1.3$-day periodicity, again proving that only the B component
participates in the periodic motion, as suggested by
\citet{Sanetal2002}.  In light of the trend in the velocities of
component A, we tried also to fit the B velocities with an additional
trend but the resulting trend was not statistically significant.

The mean radial velocity of A during the linear part is $42.5768\pm
0.0009\,\mathrm{km\,s}^{-1}$, which is very close to the
center-of-mass velocity of B -- the difference is only $\Delta RV =
0.95 \pm 0.06\,\mathrm{km\,s}^{-1}$. However, this difference depends
also on the estimated velocities we used for the templates, and
therefore we adopt a conservative error estimate of
$0.1\,\mathrm{km\,s}^{-1}$ for $\Delta RV$.

\section{Discussion}
\label{discsect}

The radial velocities derived in the previous section support the
triple-system model suggested by \citeauthor{Sanetal2002} for
\object{HD\,41004}.  Our results allow, nevertheless, a somewhat
more detailed study of the system.

\subsection{The long-term trend of the radial velocity of \object{HD\,41004\,A}}
\label{substrend}

As we have demonstrated in the previous section (see
Figure~\ref{trend}), the radial velocities of \object{HD\,41004\,A}
show a clear long-term variation over time.  For the period November
2001 to February 2002 this variation can be nicely fitted by a linear
increase.  Note, however, that one single point, at
$\mathrm{JD}=2\,451\,902.774$, which was observed one year before all
the others -- on December 2000, deviates substantially from this
linear fit.  This point may indicate that the extrapolation of the
linear approximation is no longer valid for the time of that
point. Thus, we may actually be seeing the curved part of a long-term
orbit, caused by a fourth component in the system.  Such an object
would have an orbital motion with a much shorter period than that of
the AB system, which is at least 100 years, as inferred from the
observed separation.  In fact, follow-up measurements we are currently
performing hint that this may indeed be the case, but a detailed
solution of the orbit would still be premature.

\subsection{The close orbit and the rotation of \object{HD\,41004\,Ba}}
\label{subsvsini}

As mentioned in Section~\ref{analsect}, the process by which we chose
the templates allows a very crude estimate of the equatorial
rotational velocity.  The best-fit orbit was attained when the M2
template was convolved with a rotation profile of $v\sin
i=1\,\mathrm{km\,s}^{-1}$.  Our procedure is not accurate enough to
use this value as a true measurement of $v\sin i$. However, trying
large values for $v \sin i$ degraded the quality of the orbital
solution. For example, while the quality of the solution remained
almost unchanged for $v \sin i$ as large as $5\,\mathrm{km\,s}^{-1}$,
the $O-C$~RMS was doubled at $v\sin i=10\,\mathrm{km\,s}^{-1}$.  Thus,
we believe that the equatorial velocity of Ba is of the order of a few
km\,s$^{-1}$, and probably less than 10 km\,s$^{-1}$.  Even this crude
estimate is sufficient to conclude that $v \sin i$ is substantially
lower than the expected equatorial velocity of
$\sim20\,\mathrm{km\,s}^{-1}$, assuming synchronization
\citep{Sanetal2002}.

At such a close orbit, with a period of $1.33$~days, it is usually
assumed that synchronization and alignment of the spin with the
orbital motion had been established \citep{Sanetal2002}. An obvious
explanation of the apparently small $v\sin i$ would be small
orbital inclination.  Assuming alignment, small inclination implies
that the mass of the unseen companion is considerably larger than its
derived minimum value. 

\subsection{The small eccentricity of the close orbit}
\label{subsorbev}

A binary or a star-planet system with a period as short as $1.33$~days
is na\"{\i}vely believed
to have been circularized.  However, our radial-velocity solution
shows that the close orbit has a finite non-zero eccentricity -- $e =
0.065 \pm 0.014$, which has to be explained.

Let us first estimate
the timescale of circularization due to processes occurring in Ba.
The relevant process is dissipation of the equilibrium tide through
interaction with the convective envelope \citep{Zah1989}. We follow
\citet{Rasetal1996} and write:
$$
\tau_\mathrm{circ} = \frac{\tau_\mathrm{c}}{f}
\frac{M}{M_\mathrm{env}} \frac{1}{q(1+q)}(\frac{a}{R})^8\,,
$$
where the parameter $\tau_\mathrm{c} \approx (M R^2/L)^\frac{1}{3}$ is
the eddy turnover timescale, and its numerical value is about
$0.5$\,yr.  The numerical value of $f$ is obtained by integrating the
viscous dissipation of the tidal energy through the convective zone
and is of order unity \citep{Zah1977}.  In our case the tidal pumping
period -- $P/2$ -- is much smaller than $\tau_\mathrm{c}$, and
therefore only convective eddies with turn-over time less than $P/2$
contribute to the dissipation.  The value of $f$ is then reduced to
$(P/(2\tau_\mathrm{c}))^\alpha$ \citep{Zah1989}. The correct value of
$\alpha$ is debated but generally assumed to be either $1$
\citep{Zah1992} or $2$ \citep{GolKee1977,GolMaz1991}.  We assume that
the mass of the convective envelope -- $M_\mathrm{env}$ -- is very
close to the stellar mass -- $M$ \citep[e.g.,][]{ChaBar2000}.  The
resulting timescale is about $10^{10}$\,yr, assuming $\alpha=1$, and
certainly much larger if $\alpha=2$. At the age of $1.5$\,Gyr, it
therefore seems that circularization through dissipation in Ba has not
taken place.

Now let us examine the possible involvement of Bb in circularization
processes.  If the object is substellar or maybe even a very late
M-dwarf, radiative zones in the atmosphere may form either due to the
internal physics \citep{Buretal1997} or due to the external heating by
the primary \citep{Guietal1996}. These would necessitate another
approach to calculate the circularization timescale, probably based on
dissipation of the dynamical tide \citep{Zah1977}.  The theory of
orbital evolution through tidal dissipation is still debated,
specially the mechanisms for dissipation of the dynamical tide
\citep[e.g.,][]{Claetal1995}, and therefore it is not clear whether it
contributes significantly in the case of \object{HD\,41004\,B}.

\section{Conclusion}
\label{concsect}

We have presented in this work the first application of multi-order
TODCOR to echelle spectra. The case of \object{HD\,41004} demonstrates
the unique capabilities of this technique. It utilizes all the
available prior knowledge regarding {\it both} spectral components
(the different templates), it is not limited to a fixed flux ratio (we
used the SED to calculate it) and it incorporates optimally the data
from all the relevant spectral orders.  Based on the best-fit
solutions of the present case, we estimate the precision of the
secondary velocity of \object{HD\,41004} to be
$0.56\,\mathrm{km\,s}^{-1}$ and that of the primary velocities to be
$10\,\mathrm{m\,s}^{-1}$. The radial velocities yielded accurate
orbital elements of the unseen companion of \object{HD\,41004\,B}, and
an accurate measurement of the radial acceleration of
\object{HD\,41004\,A}. The latter suggested an additional component
may be present in the system.  Follow-up observations, currently
underway, tend to confirm this hypothesis.

The combination of multi-order TODCOR together with the high
signal-to-noise and high resolution of the CORALIE spectra render this
analysis a very promising path toward expanding the database of
spectroscopic binaries and multiple systems (like \object{HD\,41004}).
It may also facilitate the detection of planets in binary stellar
systems, which are lately the focus of an increasing interest.

\begin{acknowledgements}

This research was supported by the Israeli Science Foundation (grant
no. 40/00). Support from Funda\c{c}\~{a}o para a Ci\^{e}ncia e
Tecnologia, Portugal, to N.C.S. in the form of a scholarship is
gratefully acknowledged.

\end{acknowledgements}

\bibliographystyle{aa}
\bibliography{ref}

\begin{thebibliography}{27}
\expandafter\ifx\csname natexlab\endcsname\relax\def\natexlab#1{#1}\fi

\bibitem[{{Burrows} {et~al.}(1997){Burrows}, {Marley}, {Hubbard}, {Lunine},
  {Guillot}, {Saumon}, {Freedman}, {Sudarsky}, \& {Sharp}}]{Buretal1997}
{Burrows}, A., {Marley}, M., {Hubbard}, W.~B., {et~al.} 1997, \apj, 491, 856

\bibitem[{{Chabrier} \& {Baraffe}(2000)}]{ChaBar2000}
{Chabrier}, G. \& {Baraffe}, I. 2000, \araa, 38, 337

\bibitem[{{Claret} {et~al.}(1995){Claret}, {Gimenez}, \& {Cunha}}]{Claetal1995}
{Claret}, A., {Gimenez}, A., \& {Cunha}, N.~C.~S. 1995, \aap, 299, 724

\bibitem[{{Delfosse} {et~al.}(1999){Delfosse}, {Forveille}, {Beuzit}, {Udry},
  {Mayor}, \& {Perrier}}]{Deletal1999}
{Delfosse}, X., {Forveille}, T., {Beuzit}, J.-L., {et~al.} 1999, \aap, 344, 897

\bibitem[{{Delfosse} {et~al.}(1998){Delfosse}, {Forveille}, {Perrier}, \&
  {Mayor}}]{Deletal1998}
{Delfosse}, X., {Forveille}, T., {Perrier}, C., \& {Mayor}, M. 1998, \aap, 331,
  581

\bibitem[{{Goldman} \& {Mazeh}(1991)}]{GolMaz1991}
{Goldman}, I. \& {Mazeh}, T. 1991, \apj, 376, 260

\bibitem[{{Goldreich} \& {Keeley}(1977)}]{GolKee1977}
{Goldreich}, P. \& {Keeley}, D.~A. 1977, \apj, 211, 934

\bibitem[{{Gray}(1976)}]{Gra1976}
{Gray}, D.~F. 1976, {The observation and analysis of stellar photospheres} (New
  York: Wiley-Interscience)

\bibitem[{{Guillot} {et~al.}(1996){Guillot}, {Burrows}, {Hubbard}, {Lunine}, \&
  {Saumon}}]{Guietal1996}
{Guillot}, T., {Burrows}, A., {Hubbard}, W.~B., {Lunine}, J.~I., \& {Saumon},
  D. 1996, \apjl, 459, L35

\bibitem[{{H{\" u}nsch} {et~al.}(1999){H{\" u}nsch}, {Schmitt}, {Sterzik}, \&
  {Voges}}]{Hunetal1999}
{H{\" u}nsch}, M., {Schmitt}, J.~H.~M.~M., {Sterzik}, M.~F., \& {Voges}, W.
  1999, \aaps, 135, 319

\bibitem[{{Lucy} \& {Sweeney}(1971)}]{LucSwe1971}
{Lucy}, L.~B. \& {Sweeney}, M.~A. 1971, \aj, 76, 544

\bibitem[{{Mazeh} {et~al.}(2002){Mazeh}, {Prato}, {Simon}, {Goldberg},
  {Norman}, \& {Zucker}}]{Mazetal2002}
{Mazeh}, T., {Prato}, L., {Simon}, M., {et~al.} 2002, \apj, 564, 1007

\bibitem[{{Mazeh} \& {Zucker}(1994)}]{MazZuc1994}
{Mazeh}, T. \& {Zucker}, S. 1994, \apss, 212, 349

\bibitem[{{Pickles}(1998)}]{Pic1998}
{Pickles}, A.~J. 1998, \pasp, 110, 863

\bibitem[{{Prato} {et~al.}(2002){Prato}, {Simon}, {Mazeh}, {McLean}, {Norman},
  \& {Zucker}}]{Praetal2002}
{Prato}, L., {Simon}, M., {Mazeh}, T., {et~al.} 2002, \apj, 569, 863

\bibitem[{{Queloz} {et~al.}(1999){Queloz}, {Casse}, \& {Mayor}}]{Queetal1999}
{Queloz}, D., {Casse}, M., \& {Mayor}, M. 1999, in ASP Conf. Ser. 185: IAU
  Colloq. 170: Precise Stellar Radial Velocities, 13

\bibitem[{{Rasio} {et~al.}(1996){Rasio}, {Tout}, {Lubow}, \&
  {Livio}}]{Rasetal1996}
{Rasio}, F.~A., {Tout}, C.~A., {Lubow}, S.~H., \& {Livio}, M. 1996, \apj, 470,
  1187

\bibitem[{{Santos} {et~al.}(2000){Santos}, {Israelian}, \&
  {Mayor}}]{Sanetal2000}
{Santos}, N.~C., {Israelian}, G., \& {Mayor}, M. 2000, \aap, 363, 228

\bibitem[{{Santos} {et~al.}(2001){Santos}, {Israelian}, \&
  {Mayor}}]{Sanetal2001}
---. 2001, \aap, 373, 1019

\bibitem[{{Santos} {et~al.}(2002){Santos}, {Mayor}, {Naef}, {Pepe}, {Queloz},
  {Udry}, {Burnet}, {Clausen}, {Helt}, {Olsen}, \& {Pritchard}}]{Sanetal2002}
{Santos}, N.~C., {Mayor}, M., {Naef}, D., {et~al.} 2002, \aap, 392, 215

\bibitem[{{Simkin}(1974)}]{Sim1974}
{Simkin}, S.~M. 1974, \aap, 31, 129

\bibitem[{{Tonry} \& {Davis}(1979)}]{TonDav1979}
{Tonry}, J. \& {Davis}, M. 1979, \aj, 84, 1511

\bibitem[{{Zahn}(1977)}]{Zah1977}
{Zahn}, J.-P. 1977, \aap, 57, 383

\bibitem[{{Zahn}(1989)}]{Zah1989}
---. 1989, \aap, 220, 112

\bibitem[{{Zahn}(1992)}]{Zah1992}
{Zahn}, J.~P. 1992, in Binaries as Tracers of Stellar Formation., eds. A.
  Duquennoy \& M. Mayor (Cambridge: Cambridge University Press), 253

\bibitem[{{Zucker}(2003)}]{Zuc2003}
{Zucker}, S. 2003, \mnras, {submitted}

\bibitem[{{Zucker} \& {Mazeh}(1994)}]{ZucMaz1994}
{Zucker}, S. \& {Mazeh}, T. 1994, \apj, 420, 806

\end{thebibliography}

\end{document}